\documentclass[11pt,a4paper,reqno]{amsart}

\usepackage{amsmath}
\usepackage{amssymb}
\usepackage{amsthm}
\usepackage{amsfonts}
\usepackage{amstext}
\usepackage{amsopn}
\usepackage{amsxtra}
\usepackage{graphicx}
\usepackage[latin1]{inputenc}
\usepackage{dsfont}

\newcommand\R{{\ensuremath {\mathbb R} }}
\newcommand\C{{\ensuremath {\mathbb C} }}

\renewcommand\phi{\varphi}
\newcommand{\alp}{\boldsymbol{\alpha}}
\newcommand{\alphaph}{\alpha_{\rm ph}}
\newcommand{\rhoph}{\rho_{\rm ph}}
\newcommand{\gH}{\mathfrak{H}}

\newcommand{\bA}{\boldsymbol{A}}

\newcommand{\bE}{\boldsymbol{E}}
\newcommand{\gS}{\mathfrak{S}}

\renewcommand{\to}{\rightarrow}

\newcommand{\CJ}{\mathcal{C}}
\newcommand{\cF}{\mathcal F}

\newcommand{\tr}{{\rm Tr}\,}
\newcommand{\cE}{\mathcal{E}}
\newcommand{\cC}{\mathcal{C}}
\newcommand{\cK}{\mathcal{K}}
\newcommand{\cX}{\mathcal{X}}
\newcommand\ii{{\ensuremath {\infty}}}

\newcommand{\norm}[1]{ \left| \! \left| #1 \right| \! \right| }

\newtheorem{theorem}{Theorem}
\newtheorem{remark}{Remark}

\usepackage[latin1]{inputenc}

\begin{document}

\title{Renormalization of Dirac's Polarized Vacuum}

\author{Mathieu LEWIN}

\address{CNRS \& University of Cergy-Pontoise,
95 000 Cergy-Pontoise, France}
\email{lewin@math.cnrs.fr}

\maketitle

\begin{abstract}
We review recent results on a mean-field model for relativistic electrons in atoms and molecules, which allows to describe at the same time the self-consistent behavior of the polarized Dirac sea. We quickly derive this model from Quantum Electrodynamics and state the existence of solutions, imposing an ultraviolet cut-off $\Lambda$. We then discuss the limit $\Lambda\to\ii$ in detail, by resorting to charge renormalization.



\bigskip

\noindent\scriptsize Proceedings of the Conference QMath 11 held in Hradec Kr\'alov\'e (Czechia) in September 2010. \textit{\copyright~ 2010 by the author. This paper may be reproduced, in its entirety, for non-commercial~purposes.} 

\end{abstract}

\bigskip
\bigskip

For heavy atoms, it is necessary to take relativistic effects into account. However there is no equivalent of the well-known $N$-body (non-relativistic) Schrödinger theory involving the Dirac operator, because of its negative spectrum.
The correct theory is Quantum Electrodynamics (QED). This theory has a remarkable predictive power but its
description in terms of perturbation theory restricts its range of applicability.
In fact a mathematically consistent formulation of the
nonperturbative theory is still unknown. On the other hand, effective models deduced from nonrelativistic theories (like the Dirac-Hartree-Fock model \cite{Swirles-35,EstSer-99}) suffer from inconsistencies: for instance a ground state never minimizes the physical energy which is always unbounded from below.

Here we present an effective model based on a physical energy which can be minimized to obtain the ground state in a chosen charge sector. Our model describes the behavior of a finite number of particles (electrons), coupled to that of the Dirac sea which can become polarized. Our existence results are fully non-perturbative. Like in QED, the model contains divergences which have to be removed by renormalization.

We review several results obtained in collaboration with Christian Hainzl, Philippe Gravejat, \'Eric Séré and Jan Philip Solovej. These works have already been summarized in \cite{HaiLewSerSol-07} and in the fourth chapter of \cite{Lewin-HDR}, to which the interested reader is refered for more details.

\section{A nonlinear Dirac equation}\label{sec:nonlinear_eq}
We present a mean-field model describing the self-consistent behavior of a finite number of `real' electrons in an atom or a molecule, and, simultaneously, of the infinitely many `virtual' electrons of the Dirac sea. The state of the system is described by a \emph{one-body density matrix $P$} which is a self-adjoint operator acting on the Hilbert space $\gH:=L^2(\mathbb{R}^2,\mathbb{C}^4)$, and satisfying the constraint $0\leq P\leq 1$ ($1$ denotes the identity operator on $\gH$). The operator $P$ describes the whole system consisting of the real and virtual electrons. We are interested in the following stationary equation \cite{ChaIra-89,HaiLewSerSol-07}:
\begin{equation}
\boxed{\begin{cases}
P=\chi_{(-\infty,\mu)}\left(D\right)+\delta\\
D=D^0+\alpha(\rho_{P-1/2}-\nu)\ast|x|^{-1}+X_P.
\end{cases}}
\label{eq:SCF_no_cut_off} 
\end{equation}
In this section we explain the meaning of this equation at a formal level, before turning to rigorous results.

The operator $D$ is a \emph{mean-field} Hamiltonian which is seen by all the particles. The first term 
$$D^0:=\boldsymbol{\alpha}\cdot(-i\nabla)+\beta$$
is the usual free Dirac operator \cite{Thaller} (with $\alp=(\alpha_1,\alpha_2,\alpha_3)$ where $\alpha_j$ are the usual Dirac matrices). For the sake of simplicity we have chosen units in which the speed of light is $c=1$ and the mass of the electron is $m=1$. The operator $D^0$ satisfies
$(D^0)^2=-\Delta+1$
and its spectrum is $\sigma(D^0)=(-\ii,-1]\cup[1,\ii)$.
The second term in the formula of $D$ is the Coulomb potential induced by both a fixed external density of charge $\nu$ (typically $\nu=Z\delta_0$ for a pointwise nucleus of charge $Z$ located at the origin), and the self-consistent density $\rho_{P-1/2}$ of the electrons (defined below). The number $\alpha=e^2$ which is the square of the (bare) charge $e$ of the electron, is called the \emph{bare coupling constant}. It will be renormalized later.

The third term $X_P$ in the definition of $D$ is an \emph{exchange} operator whose form depends on the chosen model. In \emph{Hartree-Fock (HF) theory} \cite{ChaIra-89,HaiLewSerSol-07}, we have
\begin{equation}
 X_P(x,y)=-\alpha\frac{(P-1/2)(x,y)}{|x-y|}
\label{eq:X_P_HF}
\end{equation}
which is called the \emph{exchange term}. In \emph{Relativistic Density Functional Theory} \cite{EngDre-87,Engel-02},
\begin{equation}
X_P=\frac{\partial F_{\text{xc}}}{\partial\rho}(\rho_{P-1/2})
\label{eq:X_P_DFT} 
\end{equation}
is the derivative of a chosen effective exchange-correlation functional, which depends only on the density $\rho_{P-1/2}$. In \emph{reduced Hartree-Fock (rHF) theory}, we simply take
$$\boxed{X_P=0.}$$
For the sake of clarity, we will mainly present the mathematical results that have been obtained in the simplest case of $X_P=0$ and we will only make comments on the Hartree-Fock case \eqref{eq:X_P_HF}. The exchange-correlation approximation \eqref{eq:X_P_DFT} has not been considered rigorously so far.

We use the notation $\chi_{(-\ii,\mu)}(D)$ to denote the spectral projector of $D$ associated with the interval $(-\ii,\mu)$. Hence Equation \eqref{eq:SCF_no_cut_off} means that the electrons of the system fill all the energies of the mean-field Hamiltonian $D$, up to the Fermi level $\mu\in(-1,1)$. In practice we choose the chemical potential $\mu$ to fix the total charge of the system.\footnote{If the external field is not too strong, fixing the charge is the same as fixing the number of electrons. However in strong fields, electron-positron pairs can be created and fixing the charge might not lead to the expected number of electrons.} 
We have added in \eqref{eq:SCF_no_cut_off} the possibility of having a density matrix $0\leq\delta\leq\chi_{\{\mu\}}(D)$ at the Fermi level, as is usually done in reduced Hartree-Fock theory \cite{Solovej-91}. So the operator $P$ is not necessarily a projector but we still use the letter $P$ for convenience. Later we will restrict ourselves to the case of $P$ being an orthogonal projector.

Equation \eqref{eq:SCF_no_cut_off} is well-known in the physical literature. A model of the same form (with an exchange term $X_P$ different from \eqref{eq:X_P_HF}) was proposed by Chaix and Iracane in \cite{ChaIra-89}. Relativistic Density Functional Theory aims at solving the same Equation \eqref{eq:SCF_no_cut_off} with $X_P$ given by \eqref{eq:X_P_DFT} and additional classical electromagnetic terms accounting for the interactions with photons, see, e.g., \cite[Eq. (6.2)]{EngDre-87} and \cite[Eq. (62)]{Engel-02}. Dirac already considered in \cite{Dirac-34b} the first order term obtained from \eqref{eq:SCF_no_cut_off} in an expansion in powers of $\alpha$, assuming $X_P=0$.

Let us now elaborate on the exact meaning of $\rho_{P-1/2}$. The charge density of an operator $A:\gH\to\gH$ with integral kernel $A(x,y)_{\sigma,\sigma'}$ is formally defined as $\rho_A(x)=\sum_{\sigma=1}^4A(x,x)_{\sigma,\sigma}=\tr_{\C^4}(A(x,x))$. In usual Hartree-Fock theory, the charge density is $\rho_P(x)$. However, as there are infinitely many particles, this does not make sense here. In \eqref{eq:SCF_no_cut_off}, the subtraction of half the identity is a convenient way to give a meaning to the density, independently of any reference, as we will explain later. One has formally, when $P$ is a projector,
$$\rho_{P-1/2}(x)=\rho_{\frac{P-P^\perp}2}(x)=\frac12\sum_{i\geq1}|\varphi_i^-(x)|^2-|\varphi_i^+(x)|^2$$
where $\{\varphi_i^-\}_{i\geq1}$ is an orthonormal basis of $P\gH$ and $\{\varphi_i^+\}_{i\geq1}$ is an orthonormal basis of $(1-P)\gH$. As was explained in \cite{HaiLewSol-07} (see also Section \ref{sec:derivation}), subtracting $1/2$ to the density matrix $P$ renders the model invariant under charge conjugation.

\subsection*{The free vacuum}

When there is no external field ($\nu\equiv0$) and when $X_P=0$, Equation \eqref{eq:SCF_no_cut_off} has an obvious solution for any $\mu\in(-1,1)$, the state made of all electrons with negative energy\footnote{In the Hartree-Fock case \eqref{eq:X_P_HF}, the free Dirac sea $P=P^0_-$ is no more a solution of \eqref{eq:SCF_no_cut_off} when $\nu\equiv0$. The Hartree-Fock free vacuum solving the nonlinear equation \eqref{eq:SCF_no_cut_off} was constructed in \cite{LieSie-00,HaiLewSol-07}, assuming an ultraviolet cut-off.}
$$P=P^0_-:=\chi_{(-\ii,0)}(D^0),$$
in accordance with Dirac's ideas \cite{Dirac-28,Dirac-30,Dirac-33}. Indeed $\rho_{P^0_--1/2}\equiv0$, as is seen by writing in the Fourier representation
$$(P^0_--1/2)(p)=-\frac{\boldsymbol{\alpha}\cdot p+\beta}{2\sqrt{1+|p|^2}}$$
and using that the Dirac matrices are trace-less. This shows the usefulness of the subtraction of half the identity to $P$, since the free vacuum $P^0_-$ now has a vanishing density.  

For a general state $P$, we can use this to write (formally):
\begin{equation}
\rho_{P-1/2}=\rho_{P-1/2}-\rho_{P^0_--1/2}=\rho_{P-P^0_-}. 
\label{eq:rho_P}
\end{equation}
When $P$ belongs to a suitable class of perturbations of $P^0_-$ (for instance when $P-P^0_-$ is locally trace-class), the density $\rho_{P-P^0_-}$ is a well-defined mathematical object. We will give below natural conditions which garantee that $P-P^0_-$ has a well-defined density in our context.

\subsection*{Electrons interacting with the polarized vacuum}

\begin{figure}
\centering
\begin{tabular}{c|c}
\includegraphics[width=6cm]{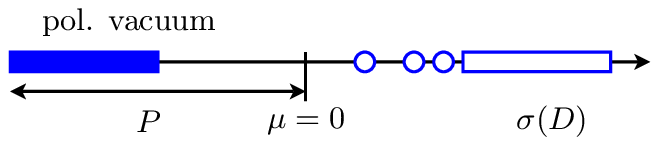}&\includegraphics[width=6cm]{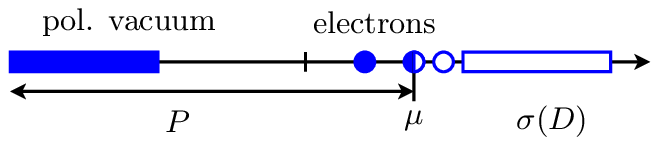}
\end{tabular}

\medskip

\caption{State of the system depending on the value of the chemical potential $\mu$.\label{fig:spectre}}
\end{figure}

With external field ($\nu\neq0$), Equation \eqref{eq:SCF_no_cut_off} models a system of electrons in the presence of a nucleus and with a self-consistent polarized Dirac sea. The number of `real' electrons in the system will depend on the value of $\mu$. Typically (for not too strong fields) when $\mu=0$, one obtains the ground state of the polarized vacuum in the presence of $\nu$, without any real electron. On the other hand $\mu>0$ in general leads to systems with a finite number of real electrons (Fig. \ref{fig:spectre}).

Note that for a generic operator $0\leq P\leq 1$ there is no natural distinction between real and virtual electrons. It is only for a solution of Equation \eqref{eq:SCF_no_cut_off} that we can interpret the eigenfunctions corresponding to the positive eigenvalues of $D$ as describing `real' electrons, and the rest of the spectrum as being the Dirac sea.

When $\mu>0$ (and $\delta=0$), the $N$ filled eigenfunctions of $D$ corresponding to the eigenvalues in $[0,\mu)$ solve the following system of nonlinear equations:
$$\left(D^0+\alpha\left(\sum_{i=1}^{N}|\phi_i|^2-\nu\right)\ast\frac{1}{|x|}+\alpha\, \rho_{P_\text{vac}-1/2}\ast\frac{1}{|x|}\right)\phi_i=\epsilon_i\,\phi_i$$
for $i=1,...,N$. This equation has the same form as the well-known Dirac-Hartree-Fock equations \cite{Swirles-35,EstSer-99}, without exchange term, and with an additional vacuum polarization potential induced by the vacuum state $P_\text{vac}:=\chi_{(-\ii,0)}(D)$. This remark was used by Chaix and Iracane in \cite{ChaIra-89} as a justification to the Dirac-Hartree-Fock model.

Most of the material of this section is purely formal and many objects (like $\rho_{P-1/2}$) do not really make sense as such. In Section \ref{sec:existence} we will introduce an ultraviolet cut-off $\Lambda$ and present rigorous results. We however first explain how the formal Equation \eqref{eq:SCF_no_cut_off} can be derived from Quantum Electrodynamics (QED).

\section{Derivation from Quantum Electrodynamics}\label{sec:derivation}
In this section we derive Equation \eqref{eq:SCF_no_cut_off} from first principles. We start with the \emph{formal} QED Hamiltonian written in Coulomb gauge, in the presence of an external density of charge $\nu$ and an external magnetic potential $a$, see \cite{Heisenberg-34,HeiEul-36,Serber-36,Schwinger-48,BjorkenDrell-65}:
\begin{multline}
\mathbb{H}^{\nu,a}= \int \Psi^*(x)\Big(\alp\cdot\big\{-i\nabla-\sqrt{\alpha}(\bA(x)+a(x)\big\}+m\beta\Big)\Psi(x) \,dx\\
-\alpha \iint \frac{\rho(x)\nu(y)}{|x-y|}dx\,dy
+ \frac{\alpha}2 \iint \frac{\rho(x)\rho(y)}{|x-y|}dx\,dy+H_f.
\label{Ham}
\end{multline}
Here $\Psi(x)$ is the second quantized field operator which annihilates an electron at $x$ and satisfies the anticommutation relation
\begin{equation}
 \Psi^*(x)_\sigma\Psi(y)_\nu+\Psi(y)_\nu\Psi^*(x)_\sigma = 2\delta_{\sigma,\nu}\delta(x-y).
\label{CAR}
\end{equation}
In the formula of $\mathbb{H}^{\nu,a}$, $\rho(x)$ is the \emph{density operator} defined by
\begin{equation}
\rho(x) =  \sum_{\sigma
=1}^4\frac{[\Psi_\sigma^*(x),\Psi_\sigma(x)]}{2}
\label{def_rho}
\end{equation}
where $[a,b]=ab-ba$. The operator $H_f$ describes the kinetic energy of the photons:
$$H_f=\frac{1}{8\pi}\int \left(|\nabla\times\bA(x)|^2+|\bE_t(x)|^2\right)\,dx=\sum_{\lambda=1,2}\int_{\R^3}dk\, |k|a^*_\lambda(k)a_\lambda(k)+\text{Cst}$$
(Cst indicates a constant which diverges in infinite volume). 
The operators $\bA(x)$ and $\bE_t(x)$ are the electromagnetic field operators for the photons and $a^*_\lambda(k)$ is the creation operator of a photon with momentum $k$ and polarization $\lambda$.
The Hamiltonian $\mathbb{H}^{\nu,a}$ formally acts on the Fock space
$\cF=\cF_{\rm e}\otimes\cF_{\rm ph}$
where $\cF_{\rm e}$ is the fermionic Fock space for the electrons and $\cF_{\rm ph}$ is the bosonic Fock space for the photons.

We emphasize that \eqref{Ham} does not contain any normal-ordering or notion of (bare) electrons and positrons: $\Psi(x)$ can annihilate electrons of negative kinetic energy. The distinction between electrons and positrons should be a result of the theory and not an input. The commutator used in the formula \eqref{def_rho} of $\rho(x)$ is a kind of renormalization, independent of any reference. It is due to Heisenberg \cite{Heisenberg-34} (see also \cite[Eq. $(96)$]{Pauli-41}) and it is necessary for a covariant formulation of QED, see \cite[Eq. $(1.14)$]{Schwinger-48} and \cite[Eq. $(38)$]{Dyson-49a}. More precisely, the Hamiltonian $\mathbb{H}^{\nu,a}$ possesses the interesting property of being invariant under charge conjugation since the following relations hold formally
$$\CJ\rho(x)\CJ^{-1}=-\rho(x),\qquad \CJ\mathbb{H}^{\nu,a} \CJ^{-1}=\mathbb{H}^{-\nu,a},$$
where $\CJ$ is the charge conjugation operator acting on the Fock space.

\medskip

We now make two approximations:
$(i)$ we neglect photons and assume there is no external magnetic field, $a\equiv0$; $(ii)$ we work in a mean-field theory, i.e. we restrict the Hamiltonian $\mathbb{H}^{\nu,a}$ to (generalized) Hartree-Fock states.

Let us recall that the electronic \emph{one-body density matrix} (two point function) of any electronic state $|\Omega\rangle\in\cF_{\rm e}$ is defined as
$$P(x,y)_{\sigma,\sigma'}=\langle\Omega|\Psi^*(x)_\sigma\Psi(y)_{\sigma'}|\Omega\rangle$$
and it satisfies $0\leq P\leq 1$.
Generalized Hartree-Fock states form a subset of (mixed) states which are completely determined by their density matrix $P$, see \cite{BacLieSol-94}. The value of any product of creation and annihilation operators is computed by means of Wick's formula.
The energy of a Hartree-Fock state $|\text{HF}\rangle\otimes|0\rangle$ (with $|0\rangle\in\cF_{\rm ph}$ being the photonic vacuum) is \cite{HaiLewSol-07}
$$\langle\mathbb{H}^{\nu,0}\rangle=\cE_{\rm HF}^\nu(P-1/2)+\text{Cst}$$
where $\text{Cst}$ is a constant (diverging in the infinite volume limit) and
\begin{multline}
 \cE_{\rm HF}^\nu(P-1/2)=\tr(D^0(P-1/2))-\alpha\iint \frac{\rho_{P-1/2}(x)\nu(y)}{|x-y|}\,dx\,dy\\
+\frac{\alpha}{2}\iint\frac{\rho_{P-1/2}(x)\rho_{P-1/2}(y)}{|x-y|}dx\,dy
-\frac{\alpha}{2}\iint\frac{|(P-1/2)(x,y)|^2}{|x-y|}dx\,dy.
\label{HF_QED}
\end{multline}

The reader can recognize in \eqref{HF_QED} the well-known Hartree-Fock energy \cite{LieSim-77,BacLieSol-94}, but applied to the ``renormalized'' density matrix $P-1/2$ instead of the usual density matrix $P$. The last two terms of the first line are respectively the kinetic energy and the interaction energy of the electrons with the external potential induced by the charge distribution $\nu$. In the second line appear respectively the so-called \emph{direct} and \emph{exchange} terms.
In Relativistic Density Functional Theory (RDFT) \cite{Engel-02, EngDre-87}, the exchange term is approximated by an exchange-correlation functional $F_{\text{xc}}(\rho_{P-1/2})$ whereas in reduced Hartree-Fock theory, the exchange term is simply dropped. 

Writing the first and second order stationarity conditions with respect to the density matrix $P$ leads to the nonlinear equation \eqref{eq:SCF_no_cut_off} with $\mu=0$. The equation with $\mu\neq0$ is obtained by replacing $D^0$ by $D^0-\mu$.
Again our derivation is formal but (in the Hartree-Fock case) this was made rigorous by means of a thermodynamic limit in \cite{HaiLewSol-07}.

\begin{remark}
Instead of the vacuum, one can take a coherent state for the photons. This leads to a classical unknown magnetic field $A(x)$ interacting with the particles. So far, there are no mathematical results on such a model.
\end{remark}

\section{Existence and non existence of solutions}\label{sec:existence}
In the presence of an external field ($\nu\neq0$), Equation \eqref{eq:SCF_no_cut_off} has \emph{no solution} in any `reasonable' Banach space \cite{HaiLewSer-05b} and it is necessary to introduce an ultraviolet regularization parameter $\Lambda$. The simplest method (although probably not optimal regarding regularity issues \cite{GraLewSer-09}) is to impose a cut-off at the level of the Hilbert space, that is to replace $\gH=L^2(\R^3;\C^4)$ by
$$\gH_\Lambda:=\{f\in L^2(\R^3;\C^4),\ {\rm supp}(\widehat{f})\subset B(0;\Lambda)\}$$
and to solve, instead of \eqref{eq:SCF_no_cut_off}, the regularized equation in $\gH_\Lambda$:
\begin{equation}
\boxed{\begin{cases}
P=\chi_{(-\infty,\mu)}\left(D\right)+\delta\\
D=\Pi_\Lambda\left(D^0+\alpha(\rho_{P-P^0_-}-\nu)\ast|x|^{-1}\right)\Pi_\Lambda
\end{cases}}
\label{eq:SCF_cut_off} 
\end{equation}
where $\Pi_\Lambda$ is the orthogonal projector onto $\mathfrak{H}_\Lambda$ in $\gH$. We take $X_P=0$ in the rest of the paper, that is we work in the reduced Hartree-Fock approximation. Note that we have used \eqref{eq:rho_P} to replace $\rho_{P-1/2}$ by $\rho_{P-P^0_-}$.

\subsection*{Existence of solutions}
Existence of solutions to \eqref{eq:SCF_cut_off} was proved in \cite{HaiLewSer-05b} for $\mu=0$ and in \cite{GraLewSer-09} for $\mu\in(-1,1)$, for all values of the coupling constant $\alpha\geq0$. The precise statement of this nonperturbative result is the following:
\begin{theorem}[Nonperturbative existence of solutions to \eqref{eq:SCF_cut_off}, \cite{HaiLewSer-05b,GraLewSer-09}]\label{thm:existence}
Assume that $\alpha\geq0$, $\Lambda>0$ and $\mu\in(-1,1)$ are given. Let $\nu$ be in the so-called \emph{Coulomb space}:
$$\mathcal{C}:=\left\{f\ :\ \int_{\R^3}|k|^{-2}|\widehat{f}(k)|^2dk<\infty\right\}.$$
Then, Equation \eqref{eq:SCF_cut_off} has at least one {solution} $P$ such that 
\begin{equation}
P-P^0_-\in\mathfrak{S}_2(\mathfrak{H}_\Lambda),\quad P^0_\pm(P-P^0_-)P^0_\pm\in\mathfrak{S}_1(\mathfrak{H}_\Lambda),\quad \rho_{P-P^0_-}\in \mathcal{C}\cap L^2(\R^3).
\label{properties}
\end{equation}
All such solutions share the same density $\rho_{P-P^0_-}$.
\end{theorem}
In \eqref{properties}, $\mathfrak{S}_1(\mathfrak{H}_\Lambda)$ and $\gS_2(\gH_\Lambda)$ are respectively the spaces of trace-class and Hilbert-Schmidt operators \cite{Simon-79} on $\mathfrak{H}_\Lambda$, and $P^0_+=1-P^0_-$.
Note that thanks to the uniqueness of $\rho_{P-P^0_-}$, the mean-field operator $D$ is also unique and only $\delta$ can differ between two solutions of \eqref{eq:SCF_cut_off}. 

Let us mention that it is natural to look for a solution of \eqref{eq:SCF_cut_off} such that $P-P^0_-$ is a Hilbert-Schmidt operator on $\mathfrak{H}_\Lambda$. If $P$ is a projector, the Shale-Stinespring theorem \cite{ShaSti-65} tells us that $P$ yields a Fock representation equivalent to that of $P^0_-$. Even when $P$ is not a projector, it will be associated with a unique Bogoliubov mixed state in the Fock space representation of $P^0_-$. This is a mathematical formulation of the statement that $P$ should not be too far from $P^0_-$. 

The method used in \cite{HaiLewSer-05b,GraLewSer-09} to prove Theorem \ref{thm:existence}, was to identify solutions of \eqref{eq:SCF_cut_off} with minimizers of the so-called \emph{reduced Bogoliubov-Dirac-Fock energy} which is nothing but the formal difference between the reduced Hartree-Fock energy of $P$ and that of the reference state $P^0_-$. A formal calculation yields with $Q=P-P^0_-$
\begin{align}
&\text{``}\,\cE_{\rm HF}^\nu(P-1/2)-\cE_{\rm HF}^0(P^0_--1/2)\,\text{''}\nonumber\\
&\ =  \tr\left(D^0Q\right) - \alpha \iint_{\R^3\times\R^3}\frac{\rho_{Q}(x)\nu(y)}{|x-y|}dx\,dy+\frac{\alpha}{2}\iint_{\R^3\times\R^3}\frac{\rho_{Q}(x)\rho_Q(y)}{|x-y|}dx\,dy\nonumber\\
&\ :=\cE^\nu_{\rm BDF}(P-P^0_-).\label{formal_BDF2}
\end{align}
The energy $\cE_{\rm BDF}^\nu$ was introduced and studied with an exchange term by Chaix and Iracane in \cite{ChaIra-89} (see also \cite{ChaIraLio-89}). An adequate mathematical formalism was then  provided by Bach, Barbaroux, Helffer and Siedentop \cite{BacBarHelSie-99} in the free case $\nu=0$, and by Hainzl, Séré and the author in \cite{HaiLewSer-05a,HaiLewSer-05b} when $\nu\neq0$.

The proof then reduces to finding the appropriate functional setting in which the minimization of the energy $\cE^\nu_\text{BDF}$ makes sense, providing a solution to \eqref{eq:SCF_cut_off}. We quickly sketch the proof for the convenience of the reader.

\begin{proof}
We want to give a clear mathematical meaning to the energy \eqref{formal_BDF2} and minimize it. Let us first consider the kinetic energy term. Noticing \cite{BacBarHelSie-99} that 
$$Q=P-P^0_-\text{ with $0\leq P\leq 1$} \quad\Longleftrightarrow \quad Q^2\leq Q^{++}-Q^{--}$$
where we have used the notation $Q^{\epsilon\epsilon'}:=P^0_\epsilon QP^0_{\epsilon'}$ with $\epsilon,\epsilon'\in\{\pm\}$,
we have (assuming all terms are well-defined),
$$\tr(D^0Q)=\tr|D^0|(Q^{++}-Q^{--})\geq \tr|D^0|Q^2.$$
Hence the kinetic energy is nonnegative. Recalling $1\leq |D^0|\leq \sqrt{1+\Lambda^2}$ on $\gH_\Lambda$, we also see that it is finite if and only if $Q^{\pm\pm}\in\gS_1(\gH_\Lambda)$ and $Q\in\gS_2(\gH_\Lambda)$.
This suggests to work in the following convex subset 
$$\cK:=\left\{Q\in\cX\; :\; Q^2\leq Q^{++}-Q^{--}\right\}$$
where $\cX$ is the Banach space
$$\cX:=\left\{ Q=Q^*\in \gS_2(\gH_\Lambda)\; :\; Q^{\pm\pm}\in \gS_1(\gH_\Lambda)\right\},$$ 
and to use the following generalized kinetic energy \cite{HaiLewSer-05a}:
$$\tr_{P^0_-}(D^0Q):=\tr|D^0|(Q^{++}-Q^{--}).$$
Using the ultraviolet cut-off $\Lambda$, it was proved in \cite[Lemma 1]{HaiLewSer-08} that the map $Q\in\cX\mapsto\rho_Q\in L^2(\R^3)\cap\cC$ is continuous. Hence the energy $\cE^\nu_\text{BDF}$ is well-defined for any state $Q\in\cK$. 

Now, when $\nu\in\cC$, we can complete the square and obtain the lower bound 
\begin{align}
\cE^\nu_\text{BDF}(Q)&=\tr_{P^0_-}(D^0Q)+\frac\alpha2\iint_{\R^3\times\R^3}\frac{(\rho_{Q}-\nu)(x)(\rho_Q-\nu)(y)}{|x-y|}dx\,dy\nonumber\\
&\qquad\qquad\qquad\qquad-\frac\alpha2\iint_{\R^3\times\R^3}\frac{\nu(x)\nu(y)}{|x-y|}dx\,dy\nonumber\\
&\geq -\frac\alpha2\iint_{\R^3\times\R^3}\frac{\nu(x)\nu(y)}{|x-y|}dx\,dy.\label{eq:bound_below}
\end{align}
This proves that $\cE^\nu_\text{BDF}$ is bounded from below. 

It is then an exercise to verify that $\cE^\nu_\text{BDF}$ is convex and lower semi-continuous on the convex set $\cK$, hence that it possesses at least one minimizer $Q=P-P^0_-$. It is a solution of \eqref{eq:SCF_cut_off} with $\mu=0$. Uniqueness of the density follows from the strict convexity of $\cE^\nu_\text{BDF}$ with respect to $\rho_Q$. To deal with the case $\mu\neq0$, one replaces $D^0$ by $D^0-\mu$.
\end{proof}

The variational argument provides solutions which \emph{a priori} only satisfy \eqref{properties} but one could think that they indeed have much better properties. As we will see in Section \ref{sec:renormalization}, this intuition is partially wrong: solutions are actually quite singular.  In particular $Q$ is in general \emph{not} trace-class, which is related to renormalization.

The property $Q^{++},Q^{--}\in\mathfrak{S}_1(\mathfrak{H}_\Lambda)$ in \eqref{properties} suggests to define the total `charge' of the system by
$$q=\tr\; (Q^{++}+Q^{--}):=\tr_{P^0_-}(Q).$$
If $Q$ is trace-class then we have $\tr(Q)=\tr_{P^0_-}(Q)$ but in general $\tr(Q)$ is not well-defined. Properties of the generalized trace $\tr_{P^0_-}$ have been provided in \cite{HaiLewSer-05a}. When $P=Q+P^0_-$ is a projector, $\tr_{P^0_-}(Q)$ is always an integer which is indeed nothing but the relative index of the pair $(P,P^0_-)$, see \cite{HaiLewSer-05a,AvrSeiSim-94}. 

\begin{figure}
\centering
\includegraphics[width=10cm]{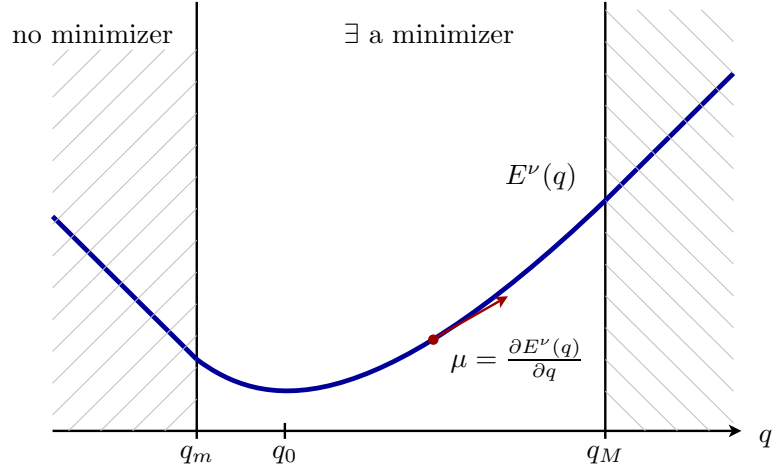}
\caption{Varying the chemical potential $\mu$ is, by convexity, equivalent to varying the total charge $q=\tr_{P^0_-}(Q)$ of the system.\label{fig:convex}}
\end{figure}

Varying $\mu$ allows to pick the desired total charge, as we now explain. Let us introduce the following constrained minimization problem
$$E^\nu(q):=\inf_{\substack{Q\in\cK\\ \tr_{P^0_-}(Q)=q}}\cE^\nu_\text{BDF}(Q).$$
The function $q\mapsto E^\nu(q)$ is convex. Assume that $Q=P-P^0_-$ is a ground state for $E^\nu_\text{BDF}(q)$. Then simple convexity arguments show that $Q$ is also a \emph{global} minimizer of the free energy $\cE^\nu_\text{BDF}-\mu\tr_{P^0_-}$, with $\mu=\partial E^\nu(q)/\partial q$. Indeed it was shown in \cite{GraLewSer-09} that $E^\nu_\text{BDF}$ is strictly convex on some interval $(q_m,q_M)$ (corresponding to $\mu\in(-1,1)$), which is also the largest interval on which $E^\nu(q)$ has ground states, see Fig. \ref{fig:convex}. Therefore varying $\mu$ in $(-1,1)$ is exactly the same as solving the minimization problem $E^\nu(q)$ for $q\in(q_m,q_M)$.

\subsection*{Maximum ionization}

The numbers $q_m$ and $q_M$ can be interpreted as the minimal and maximal possible ionization of the system in the presence of the external field $\nu$. It is important to derive bounds on these quantities, in order to determine for which values of the charge $q$ the system is stable.
The following was proved in \cite{GraLewSer-09}:

\begin{theorem}[Maximum Ionization \cite{GraLewSer-09}]\label{thm:ionization}
We assume that $\nu\in\cC\cap L^1(\R^3)$, with $Z:=\int\nu\geq0$.

\smallskip

\noindent$\bullet$ \emph{(Existence of neutral atoms)} One has $Z\in [q_m,q_M]$.

\smallskip

\noindent$\bullet$ \emph{(Ionization bounds for weak external fields)} For a regular ultraviolet cut-off, there exists constants $C,C'>0$ such that the following holds: For $\alpha\norm{\nu}_\cC+\alpha(1+\log\Lambda)\leq C'$ and $Z\geq0$, one has
\begin{multline}
-C\frac{\alpha\log\Lambda+1/\Lambda+\alpha \norm{\nu}_\cC}{1-C\alpha\log\Lambda}\leq  q_m \leq0\leq \\
\leq {Z}\leq q_M\leq \frac{{2Z}+C\left(\alpha\log\Lambda+1/\Lambda+\alpha \norm{\nu}_\cC\right)}{1-C\alpha\log\Lambda}. 
\label{eq:ionization_bound}
\end{multline}
\end{theorem}

The bound \eqref{eq:ionization_bound} is a generalization to the BDF model of an estimate due to Lieb \cite{Lieb-84}. In the nonrelativistic limit \eqref{eq:ionization_bound} reduces to Lieb's result $Z\leq q_M\leq 2Z$.

In \cite{GraLewSer-09}, the bound \eqref{eq:ionization_bound} is shown by using a more regular ultraviolet cut-off. To be more precise, the free Dirac operator $D^0$ was replaced by an operator which grows faster at infinity
$$\tilde{D}^0=D^0\left(1-\frac{\Delta}{\Lambda^2}\right)$$
and the model was settled in the whole space $\gH=L^2(\R^3,\C^4)$.

\section{Renormalization}\label{sec:renormalization}
We have described results dealing with existence (and non existence for $q\notin[q_m,q_M]$) of solutions to Equation \eqref{eq:SCF_cut_off}. All these solutions depend on the ultraviolet cut-off $\Lambda$ and it is a natural question to investigate how. Indeed, the limit $\Lambda\to\ii$ for $\alpha$ fixed was shown to be very singular in \cite[Theorem 2]{HaiLewSer-05b} and the correct way to tackle this issue is to resort to charge renormalization.

Let us start by recalling the spirit of renormalization. A physical theory usually aims at predicting physical observables in terms of the parameters in the model. Sometimes, interesting quantities are divergent and it is necessary to introduce cut-offs. In our case the parameters are the coupling constant $\alpha=e^2$, the cut-off $\Lambda$, the chemical potential $\mu$ and the external density $\nu$. For simplicity we will take $\mu=0$ and we will not emphasize the dependence in $\nu$ in our notation. In our system of units the mass of the electron is $m=1$. 
Predicted physical quantities are functions $F(\alpha,\Lambda)$. The charge $e$ (or equivalently its square, the coupling constant $\alpha$) is also a physical observable and renormalization occurs when the value predicted by the theory is different from its `bare' value:
\begin{equation}
\alphaph=\alphaph(\alpha,\Lambda)\neq \alpha.
\label{eq:ph_mass_and_charge} 
\end{equation}
In this case the parameter $\alpha$ is not observable in contrast with $\alphaph=\alphaph(m,\alpha,\Lambda)$ which has to be set equal to its experimental value. The relation \eqref{eq:ph_mass_and_charge} has to be inverted, in order to express the bare parameter in terms of the physical one:
\begin{equation}
\alpha=\alpha(\alphaph,\Lambda).
\label{eq:bare_mass_and_charge}
\end{equation}
This allows to express any observable quantity $F$ as a function $\tilde{F}$ of the physical parameters and the cut-off $\Lambda$:
\begin{equation}
\tilde F(\alphaph,\Lambda)=F\big(\alpha(\alphaph,\Lambda)\,,\,\Lambda\big)
\label{eq:ph_quantities}
\end{equation}
A possible definition of renormalizability is that \emph{all such observable quantities have a limit when $\Lambda\to\ii$, for fixed $\alphaph$}.

Important difficulties can be encountered when trying to complete this program. For instance the physical quantity $\alphaph$ might be a \emph{nonexplicit function of $\alpha$}. The corresponding formulas can then only be inverted perturbatively to any order, as is the case in QED \cite{Dyson-49b,BjorkenDrell-65,ItzyksonZuber}. 

\subsection*{Nonperturbative charge renormalization formula}
For the model presented in this article it is fortunate that there is an explicit and nonperturbative relation between $\alphaph$ and $\alpha$, as expressed in the following result:

\begin{theorem}[Nonperturbative charge renormalization formula \cite{GraLewSer-09}]\label{thm:renorm}
Assume that $\alpha\geq0$, $\Lambda>0$ and $\mu\in(-1,1)$ are given and let $P$ be a solution of \eqref{eq:SCF_cut_off} as given by Theorem \ref{thm:existence}. If $\nu\in\cC\cap L^1(\R^3)$, then $\rho_{P-P^0_-}\in L^1(\R^3)$ and it holds 
\begin{equation}
\boxed{\int_{\R^3}\nu -\int_{\R^3}\rho_{P-P^0_-}=\frac{\displaystyle\int_{\R^3}\nu-\tr_{P^0_-}(P-P^0_-)}{1+\alpha B_\Lambda}}
\label{relation_charge}
\end{equation}
where
\begin{equation}
B_\Lambda  = \frac1\pi\int_{0}^{\frac{\Lambda}{\sqrt{1+\Lambda^2}}}\frac{z^2-z^4/3}{1-z^2}dz=\frac{2}{3\pi}\log\Lambda-\frac{5}{9\pi}+\frac{2\log2}{3\pi}+O(1/\Lambda^2).
\label{expression_B}
\end{equation}
\end{theorem}

Note that, except in the neutral case $\tr_{P^0_-}(P-P^0_-)=Z$, \eqref{relation_charge} implies that the solution $Q=P-P^0_-$ found in Theorem \ref{thm:existence} cannot be trace-class. If $Q$ were trace-class, we would have $\tr_{P^0_-}(Q)=\tr(Q)=\int_{\R^3}\rho_Q$ which contradicts \eqref{relation_charge}.

The previous result is interpreted as follows. Assume that we put a nucleus of charge $Z$ in the vacuum, and let $P$ be the corresponding Dirac's polarized vacuum (that is we take $\mu=0$ in \eqref{eq:SCF_cut_off}). When $\alpha\norm{\nu}_\cC$ is small enough,\footnote{By scaling we can keep $\int_{\R^3}\nu$ fixed and choose $\alpha\norm{\nu}_\cC$ as small as we want.} it was proved in \cite{HaiLewSer-05b} that it holds $\|P-P^0_-\|<1$. This itself implies that the relative index vanishes, $\tr_{P^0_-}(P-P^0_-)=0$, hence
$$Z-\int_{\R^3}\rho_{P-P^0_-}=\frac{Z}{1+\alpha B_\Lambda}.$$
In reality we never measure the charge of the nucleus alone, but we always also observe the corresponding vacuum polarization. Hence the physical coupling constant is given by the renormalization formula
\begin{equation}
\boxed{\alpha_{\rm ph}=\frac\alpha{1+\alpha B_\Lambda} \Longleftrightarrow \alpha=\frac\alphaph{1-\alphaph B_\Lambda}.}
\label{charge_renormalization}
\end{equation}
In our theory we must fix $\alpha_{\rm ph}$ and not $\alpha$. Using \eqref{charge_renormalization} we can express any physical quantity in terms of $\alphaph$ and $\Lambda$ only. 

Unfortunately it holds $\alpha_{\rm ph} B_\Lambda<1$ hence it makes no sense to take $\Lambda\to\infty$ while keeping $\alpha_{\rm ph}$ fixed (this is the so-called Landau pole \cite{Landau-55}) and one has to look for a weaker definition of renormalizability.
The cut-off $\Lambda$ which was first introduced as a mathematical trick to regularize the model has actually a physical meaning. A natural scale occurs beyond which the model does not make sense. Fortunately, this corresponds to momenta of the order $e^{3\pi/2\alphaph}$, a huge number for $\alphaph\simeq1/137$.

\subsection*{Asymptotic renormalization}

It is convenient to define a renormalized density $\rho_{\rm ph}$ by the relation \cite{HaiLewSer-05b}
\begin{equation}
\alpha_{\rm ph}\rho_{\rm ph}=\alpha\big(\nu-\rho_{P-P^0_-}\big)
\label{eq:def_rho_ph}
\end{equation}
in such a way that $D=D^0-\alpha_{\rm ph}\rho_{\rm ph}\ast|x|^{-1}$. This procedure is similar to wavefunction renormalization. By uniqueness of $\rho_{P-P^0_-}$ we can see $\rho_{\rm ph}$ as a function of $\alpha_{\rm ph}$, $\nu$, $\mu$ and $\Lambda$ (or $\kappa$). For the sake of clarity we do not emphasize the dependence in $\nu$ and we take $\mu=0$ (this means that we consider the vacuum polarization in the presence of the nucleus, without any real electron). The self-consistent equation for $\rhoph$ was derived in \cite{HaiLewSer-05b}.

It is explained in \cite{GraLewSer-10} that one can expand $\rho_{\rm ph}=\rho_{\rm ph}(\alpha_{\rm ph},\Lambda)$ as follows:
\begin{equation}
\rho_{\rm ph}(\alpha_{\rm ph},\Lambda)=\sum_{n=0}^\infty (\alpha_{\rm ph})^n\nu_{n,\Lambda}
\label{series_cut_off}
\end{equation}
where $\{\nu_{n,\Lambda}\}_n\subset L^2(\R^3)\cap\cC$ is a sequence depending only on the external density $\nu$ and the cut-off $\Lambda$. This sequence is defined by an explicit induction formula which is detailed in \cite{GraLewSer-10} and that we do not write here for shortness. The series \eqref{series_cut_off} has a positive  radius of convergence, which is however believed to shrink to zero when $\Lambda\to\ii$.

Assuming $\widehat\nu$ decays fast enough (see condition \eqref{condition_nu}), it is proved in \cite{GraLewSer-10} that for any fixed $n$, the limit  $\nu_{n,\Lambda}\to\nu_{n}$ exists in $L^2(\R^3)\cap\mathcal{C}$. This is what is usually meant by renormalizability in QED: each term of the perturbation series in powers of the physical $\alpha_{\rm ph}$ has a limit when the cut-off is removed. The sequence $\{\nu_n\}_n$ is the one which is calculated in practice \cite{BjorkenDrell-65,GreMulRaf-85,EngDre-87,Engel-02}. One has for instance $\nu_0=\nu$ and 
$$\nu_1\ast|x|^{-1}=\frac{1}{3\pi}\int_1^\infty dt\, (t^2-1)^{1/2}\left[\frac2{t^2}+\frac{1}{t^4}\right]\int_{\R^3} e^{-2|x-y|t}\frac{\nu(y)}{|x-y|}\, dy,$$
the \emph{Uehling potential} \cite{Uehling-35,Serber-35}. All the others $\nu_n$ can be calculated by induction in terms of $\nu_0,...,\nu_{n-1}$. An explicit recursion relation is provided in \cite{GraLewSer-10}.

The next natural question is to understand the link between the well-defined, cut-off dependent, series \eqref{series_cut_off} and the \emph{formal series} $\sum_{n=0}^\infty(\alpha_{\rm ph})^n\nu_{n}$. Recall that $\alpha_{\rm ph}B_\Lambda<1$ by construction, so it is in principle not allowed to take the limit $\Lambda\to\infty$ while keeping $\alpha_{\rm ph}$ fixed. 

It is more convenient to change variables and take as new parameters $\alphaph$ and $\kappa=\alphaph B_\Lambda$, with the additional constraint that $0<\kappa<1$. The new parameter $\kappa$ is now independent of $\alphaph$ and we ask ourselves whether predicted physical quantities will depend very much on the chosen value of $0<\kappa<1$. The purpose of \cite{GraLewSer-10} was precisely to prove that the asymptotics of any physical quantity in the regime $\alphaph\ll1$ is actually \emph{independent of $\kappa$} to any order in $\alphaph$, which is what was called \emph{asymptotic renormalizability}. Note that fixing $\kappa\in(0,1)$ amounts to take an $\alphaph$-dependent cut-off $\Lambda\simeq Ce^{3\pi\kappa/2\alphaph}\gg1$.

\begin{theorem}[Asymptotic renormalization of the nuclear density \cite{GraLewSer-10}]
\label{thm:asymptotics}
Consider a function $\nu \in L^2(\R^3) \cap \cC$ such that
\begin{equation}
\label{condition_nu}
\int_{\R^3} \log (1 + |k|)^{2 N + 2} |\widehat\nu(k)|^2 dk < \ii
\end{equation}
for some integer $N$. Let $\rhoph(\alphaph,\kappa)$ be the unique physical density defined by \eqref{eq:def_rho_ph} with $\mu = 0$, $\alphaph>0$ and $0<\kappa<1$, corresponding to the bare coupling constant $\alpha= (1-\kappa)^{-1}\alphaph$ and the ultraviolet cut-off $\Lambda$ such that $B_\Lambda=\kappa/\alphaph$.

Then, for every $0<\epsilon<1$, there exist two constants $C(N, \epsilon, \nu)$ and $a(N, \epsilon, \nu)$, depending only on $N$, $\epsilon$ and $\nu$, such that one has
\begin{equation}
\label{eq:asymptotics}
\boxed{\norm{\rhoph(\alphaph, \kappa) - \sum_{n = 0}^N \nu_n (\alphaph)^n}_{L^2(\R^3) \cap \cC} \leq C(N, \epsilon, \nu)\; \alphaph^{N + 1}}
\end{equation}
for all $0 \leq \alphaph \leq a(N, \epsilon, \nu)$ and all $\epsilon \leq \kappa \leq 1 - \epsilon$.
\end{theorem}

The interpretation of Theorem \ref{thm:asymptotics} is that the renormalized density $\rhoph(\alphaph,\kappa)$ is \emph{asymptotically} given by the formal series $\sum_{n\geq0}(\alphaph)^n\nu_n$, \emph{uniformly in the renormalization parameter $\kappa$} in the range $\epsilon\leq \kappa\leq 1-\epsilon$. For a very large range of cut-offs, 
$$C_1e^{{3\epsilon\pi}/{2\alpha_{\rm ph}}}\leq \Lambda\leq C_2e^{{3(1-\epsilon)\pi}/{2\alpha_{\rm ph}}}$$
the result is independent of $\Lambda$ for small $\alphaph$. This formulation of renormalizability is more precise than the requirement that each $\nu_{n,\Lambda}$ converges. It also leads to the formal perturbation series in a very natural way. 

It was argued by Dyson in \cite{Dyson-52} that the perturbation series $\sum_{n\geq0}(\alphaph)^n\nu_n$ it is probably \emph{divergent}, but there is no mathematical proof so far. In \cite{GraLewSer-10}, some properties of the sequence $\{\nu_n\}$ were derived.

\section*{Conclusion}
We have presented a mean-field theory for electrons in atoms and molecules, which describes at the same time the self-consistent behavior of Dirac's polarized vacuum. The model can be deduced from Quantum Electrodynamics by restricting to Hartree-Fock states and neglecting photons in the Coulomb gauge.

Existence of ground states could be established, with or without a charge constraint. The so-obtained states are rather singular, in particular they yield a perturbation $Q$ of the free vacuum $P^0_-$ which is in general not trace-class but still has $\rho_Q\in L^1(\R^3)$. This technical issue is at the origin of charge renormalization.

The formula linking the physical coupling constant $\alphaph$ and the bare $\alpha$ is explicit and exhibits a Landau pole, rendering impossible to remove the ultraviolet cut-off $\Lambda$ while keeping $\alphaph$ fixed. Nevertheless in a regime where $\alphaph\ll1$ and $\Lambda\gg1$ such that $\kappa=(2/3\pi)\alphaph\log\Lambda$ stays bounded, the asymptotics is found to be independent of the value of $\kappa$, to any order in the physical coupling constant $\alphaph$. The terms of the asymptotic expansion are the ones which are computed in practice. The first order term induces the famous Uehling potential.

The model which we have presented in this paper is probably not quantitative but it already possesses several of the qualitative properties of full Quantum Electrodynamics, with the advantage that they can be studied in a fully rigorous manner. A more quantitative model would include photons, for instance via an additional self-consistent classical magnetic field, as is done in Relativistic Density Functional Theory.


\begin{thebibliography}{10}

\bibitem{AvrSeiSim-94}
{\sc J.~Avron, R.~Seiler, and B.~Simon}, {\em The index of a pair of
  projections}, J. Funct. Anal., 120 (1994), pp.~220--237.

\bibitem{BacBarHelSie-99}
{\sc V.~Bach, J.~M. Barbaroux, B.~Helffer, and H.~Siedentop}, {\em On the
  stability of the relativistic electron-positron field}, Commun. Math. Phys.,
  201 (1999), pp.~445--460.

\bibitem{BacLieSol-94}
{\sc V.~Bach, E.~H. Lieb, and J.~P. Solovej}, {\em Generalized {H}artree-{F}ock
  theory and the {H}ubbard model}, J. Statist. Phys., 76 (1994), pp.~3--89.

\bibitem{BjorkenDrell-65}
{\sc J.~D. Bjorken and S.~D. Drell}, {\em Relativistic quantum fields},
  McGraw-Hill Book Co., New York, 1965.

\bibitem{ChaIra-89}
{\sc P.~Chaix and D.~Iracane}, {\em From quantum electrodynamics to mean field
  theory: {I}. {T}he {B}ogoliubov-{D}irac-{F}ock formalism}, J. Phys. B, 22
  (1989), pp.~3791--3814.

\bibitem{ChaIraLio-89}
{\sc P.~Chaix, D.~Iracane, and P.-L. Lions}, {\em From quantum electrodynamics
  to mean field theory: {I}{I}. {V}ariational stability of the vacuum of
  quantum electrodynamics in the mean-field approximation}, J. Phys. B, 22
  (1989), pp.~3815--3828.

\bibitem{Dirac-28}
{\sc P.~A. Dirac}, {\em {The quantum theory of the electron. II}}, Proceedings
  Royal Soc. London (A), 118 (1928), pp.~351--361.

\bibitem{Dirac-30}
\leavevmode\vrule height 2pt depth -1.6pt width 23pt, {\em {A theory of
  electrons and protons}}, Proceedings Royal Soc. London (A), 126 (1930),
  pp.~360--365.

\bibitem{Dirac-33}
\leavevmode\vrule height 2pt depth -1.6pt width 23pt, {\em Theory of electrons
  and positrons}.
\newblock Nobel lecture delivered at Stockholm, 1933.

\bibitem{Dirac-34b}
\leavevmode\vrule height 2pt depth -1.6pt width 23pt, {\em Th{\'e}orie du
  positron}, Solvay report, XXV (1934), pp.~203--212.

\bibitem{Dyson-49a}
{\sc F.~J. Dyson}, {\em The radiation theories of {T}omonaga, {S}chwinger, and
  {F}eynman}, Phys. Rev. (2), 75 (1949), pp.~486--502.

\bibitem{Dyson-49b}
\leavevmode\vrule height 2pt depth -1.6pt width 23pt, {\em The {$S$} matrix in
  quantum electrodynamics}, Phys. Rev. (2), 75 (1949), pp.~1736--1755.

\bibitem{Dyson-52}
\leavevmode\vrule height 2pt depth -1.6pt width 23pt, {\em Divergence of
  {P}erturbation {T}heory in {Q}uantum {E}lectrodynamics}, Phys. Rev., 85
  (1952), pp.~631--632.

\bibitem{Engel-02}
{\sc E.~Engel}, {\em Relativistic Density Functional Theory: Foundations and
  Basic Formalism}, vol.~`Relativistic Electronic Structure Theory, Part 1.
  Fundamentals', Elsevier (Amsterdam), {S}chwerdtfeger~ed., 2002, ch.~10,
  pp.~524--624.

\bibitem{EngDre-87}
{\sc E.~Engel and R.~M. {Dreizler}}, {\em {Field-theoretical approach to a
  relativistic Thomas-Fermi-Dirac-Weizs{\"a}cker model}}, Phys. Rev. A, 35
  (1987), pp.~3607--3618.

\bibitem{EstSer-99}
{\sc M.~J. Esteban and {\'E}.~S{\'e}r{\'e}}, {\em Solutions of the
  {D}irac-{F}ock equations for atoms and molecules}, Commun. Math. Phys., 203
  (1999), pp.~499--530.

\bibitem{GraLewSer-09}
{\sc P.~Gravejat, M.~Lewin, and {\'E}.~S{\'e}r{\'e}}, {\em Ground state and
  charge renormalization in a nonlinear model of relativistic atoms}, Commun.
  Math. Phys., 286 (2009), pp.~179--215.

\bibitem{GraLewSer-10}
\leavevmode\vrule height 2pt depth -1.6pt width 23pt, {\em Renormalization and
  asymptotic expansion of {D}irac's polarized vacuum}, 2010.

\bibitem{GreMulRaf-85}
{\sc W.~Greiner, B.~M{\"u}ller, and J.~Rafelski}, {\em Quantum Electrodynamics
  of Strong Fields}, Texts and Monographs in Physics, Springer-Verlag,
  first~ed., 1985.

\bibitem{HaiLewSer-05a}
{\sc C.~Hainzl, M.~Lewin, and {\'E}.~S{\'e}r{\'e}}, {\em Existence of a stable
  polarized vacuum in the {B}ogoliubov-{D}irac-{F}ock approximation}, Commun.
  Math. Phys., 257 (2005), pp.~515--562.

\bibitem{HaiLewSer-05b}
\leavevmode\vrule height 2pt depth -1.6pt width 23pt, {\em Self-consistent
  solution for the polarized vacuum in a no-photon {QED} model}, J. Phys. A, 38
  (2005), pp.~4483--4499.

\bibitem{HaiLewSer-08}
\leavevmode\vrule height 2pt depth -1.6pt width 23pt, {\em Existence of atoms
  and molecules in the mean-field approximation of no-photon quantum
  electrodynamics}, Arch. Ration. Mech. Anal., 192 (2009), pp.~453--499.

\bibitem{HaiLewSerSol-07}
{\sc C.~Hainzl, M.~Lewin, {\'E}.~S{\'e}r{\'e}, and J.~P. Solovej}, {\em A
  minimization method for relativistic electrons in a mean-field approximation
  of quantum electrodynamics}, Phys. Rev. A, 76 (2007), p.~052104.

\bibitem{HaiLewSol-07}
{\sc C.~Hainzl, M.~Lewin, and J.~P. Solovej}, {\em The mean-field approximation
  in quantum electrodynamics: the no-photon case}, Comm. Pure Appl. Math., 60
  (2007), pp.~546--596.

\bibitem{Heisenberg-34}
{\sc W.~Heisenberg}, {\em Bemerkungen zur {D}iracschen {T}heorie des
  {P}ositrons}, Z. Phys., 90 (1934), pp.~209--231.

\bibitem{HeiEul-36}
{\sc W.~Heisenberg and H.~{Euler}}, {\em {Folgerungen aus der Diracschen
  Theorie des Positrons}}, Zeitschrift fur Physik, 98 (1936), pp.~714--732.

\bibitem{ItzyksonZuber}
{\sc C.~Itzykson and J.~B. Zuber}, {\em Quantum field theory}, McGraw-Hill
  International Book Co., New York, 1980.
\newblock International Series in Pure and Applied Physics.

\bibitem{Landau-55}
{\sc L.~D. Landau}, {\em On the quantum theory of fields}, in Niels Bohr and
  the development of physics, McGraw-Hill Book Co., New York, N. Y., 1955,
  pp.~52--69.

\bibitem{Lewin-HDR}
{\sc M.~Lewin}, {\em {L}arge {Q}uantum {S}ystems: a {M}athematical and
  {N}umerical {P}erspective.}
\newblock Habilitation {\`a} Diriger des Recherches, University of
  Cergy-Pontoise, June 2010.

\bibitem{Lieb-84}
{\sc E.~H. Lieb}, {\em Bound on the maximum negative ionization of atoms and
  molecules}, Phys. Rev. A, 29 (1984), pp.~3018--3028.

\bibitem{LieSie-00}
{\sc E.~H. Lieb and H.~Siedentop}, {\em Renormalization of the regularized
  relativistic electron-positron field}, Commun. Math. Phys., 213 (2000),
  pp.~673--683.

\bibitem{LieSim-77}
{\sc E.~H. Lieb and B.~Simon}, {\em The {H}artree-{F}ock theory for {C}oulomb
  systems}, Commun. Math. Phys., 53 (1977), pp.~185--194.

\bibitem{Pauli-41}
{\sc W.~Pauli}, {\em {Relativistic field theories of elementary particles}},
  Rev. Modern Physics, 13 (1941), pp.~203--232.

\bibitem{Schwinger-48}
{\sc J.~Schwinger}, {\em Quantum electrodynamics. {I}. {A} covariant
  formulation}, Phys. Rev. (2), 74 (1948), pp.~1439--1461.

\bibitem{Serber-35}
{\sc R.~Serber}, {\em Linear modifications in the {M}axwell field equations},
  Phys. Rev. (2), 48 (1935), pp.~49--54.

\bibitem{Serber-36}
{\sc R.~{Serber}}, {\em {A Note on Positron Theory and Proper Energies}},
  Physical Review, 49 (1936), pp.~545--550.

\bibitem{ShaSti-65}
{\sc D.~Shale and W.~F. Stinespring}, {\em Spinor representations of infinite
  orthogonal groups}, J. Math. Mech., 14 (1965), pp.~315--322.

\bibitem{Simon-79}
{\sc B.~Simon}, {\em Trace ideals and their applications}, vol.~35 of London
  Mathematical Society Lecture Note Series, Cambridge University Press,
  Cambridge, 1979.

\bibitem{Solovej-91}
{\sc J.~P. Solovej}, {\em {Proof of the ionization conjecture in a reduced
  Hartree-Fock model.}}, Invent. Math., 104 (1991), pp.~291--311.

\bibitem{Swirles-35}
{\sc B.~Swirles}, {\em {The relativistic self-consistent field.}}, Proc. R.
  Soc. Lond., Ser. A, 152 (1935), pp.~625--649.

\bibitem{Thaller}
{\sc B.~Thaller}, {\em The {D}irac equation}, Texts and Monographs in Physics,
  Springer-Verlag, Berlin, 1992.

\bibitem{Uehling-35}
{\sc E.~Uehling}, {\em Polarization effects in the positron theory}, Phys. Rev.
  (2), 48 (1935), pp.~55--63.

\end{thebibliography}

\end{document}